# Imaging and controlling vortex dynamics in mesoscopic superconductor-normal-metal-superconductor arrays

Tyler R. Naibert[1], Hryhoriy Polshyn[1,2,*], Rita Garrido-Menacho[1], Malcolm Durkin[1], Brian Wolin[1], Victor Chua[1], Ian Mondragon-Shem[1], Taylor Hughes[1], Nadya Mason[1,*], and Raffi Budakian[1,3]

[1]*Department of Physics, University of Illinois at Urbana-Champaign, 1110 W. Green St., Urbana, IL 61801-3080, USA*

[2]*Department of Physics, University of California, Santa Barbara, CA 93106, USA*

[3]*Institute for Quantum Computing, University of Waterloo, Waterloo, ON, Canada, N2L3G1*
*Department of Physics, University of Waterloo, Waterloo, ON, Canada, N2L3G1*
*Perimeter Institute for Theoretical Physics, Waterloo, ON, Canada, N2L2Y5*
*Canadian Institute for Advanced Research, Toronto, ON, Canada, M5G1Z8*

*Corresponding authors. Email: polshyn@ucsb.edu; nadya@illinois.edu

Harnessing the properties of vortices in superconductors is crucial for fundamental science and technological applications; thus, it has been an ongoing goal to locally probe and control vortices. Here, we use a scanning probe technique that enables studies of vortex dynamics in superconducting systems by leveraging the resonant behavior of a raster-scanned, magnetic-tipped cantilever. This experimental setup allows us to image and control vortices, as well as extract key energy scales of the vortex interactions. Applying this technique to lattices of superconductor island arrays on a metal, we obtain a variety of striking spatial patterns that encode information about the energy landscape for vortices in the system. We interpret these patterns in terms of local vortex dynamics and extract the relative strengths of the characteristic energy scales in the system, such as the vortex-magnetic field and vortex-vortex interaction strengths, as well as the vortex chemical potential. We also demonstrate that the relative strengths of the interactions can be tuned and show how these interactions shift with an applied bias. The high degree of tunability and local nature of such vortex imaging and control not only enable new understanding of vortex interactions, but also have potential applications in more complex systems such as those relevant to quantum computing.

## I. INTRODUCTION

Many electronic and magnetic properties of superconductors can be understood through study of the vortices present in the superconductor. Each vortex allows one flux quantum to penetrate the superconducting surface, allowing the bulk of the superconductor to remain in the superconducting state. When an electrical current is applied, vortices move transversely to the current, a dissipative process that removes the perfect electrical conductivity of the system. Forces that prevent motion of vortices, known as pinning forces, are desirable to prevent this dissipation, and to control the positions that vortices occupy. Beyond enhancing superconductivity, it is desirable to control vortex positions to predict vortex paths, as well as the fields surrounding the superconductor. The demonstration of locally probed and manipulated vortices [1-14] is relevant to a variety of technological applications, including quantum computation [15, 16].

In previous works, imaging techniques such as scanning SQUID microscopy [1-3, 17], Hall probe microscopy [18, 19], scanning tunneling microscopy [6, 20], NV center magnetometry [21, 22], and cantilever-based techniques [23-26] have played a central role in studies of vortex lattices and the internal structure of individual vortices. However, these studies focused on probing and manipulating individual vortices rather than capturing and controlling the dynamics of an entire vortex ensemble. This allowed determination of the pinning strength but not the other important energy scales such as vortex-vortex interaction strength. Yet determining vortex interactions is crucial for enabling predictive vortex manipulation and control for applications. In this article we use a technique that overcomes this obstacle: a method we term $\Phi_0$-Magnetic Force Microscopy ($\Phi_0$-MFM) [27, 28], which probes the dynamic motion of a small group of vortices (from 1 to ~ 12) trapped in the magnetic field generated by the tip of a vertically-oriented cantilever. Here, we use this technique to determine the vortex number in a pinned vortex configuration, extract the relative vortex-vortex and vortex-field interactions strengths as well as the vortex chemical potential, and probe the response to an applied current bias.

## II. $\Phi_0$-MAGNETIC FORCE MICROSCOPY OF VORTEX DYNAMICS IN SUPERCONDUCTOR-NORMAL-SUPERCONDUCTOR ARRAYS

$\Phi_0$-MFM is demonstrated on triangular arrays of Nb islands deposited on Au films [Fig. 1(a)], which form a superconductor-normal metal-superconductor (SNS) array. A resistance vs. temperature measurement showing a superconducting transition for a representative array is shown in Fig. 1(c). In this figure, $T_1$ represents the superconducting transition temperature of the Nb islands and $T_2$ the temperature at which the Au film regions between the Nb islands – the interstitial regions – become superconducting [31] and strongly Type II in behavior. Vortices will preferentially stay close to the magnetic tip but avoid the Nb islands and thus reside in the interstitial regions, since these regions host a weaker superconducting condensate. As a result, the SNS array system of superconducting islands on top of normal metal films proves to be a controllable and tunable model for superconducting films having a periodic pinning potential

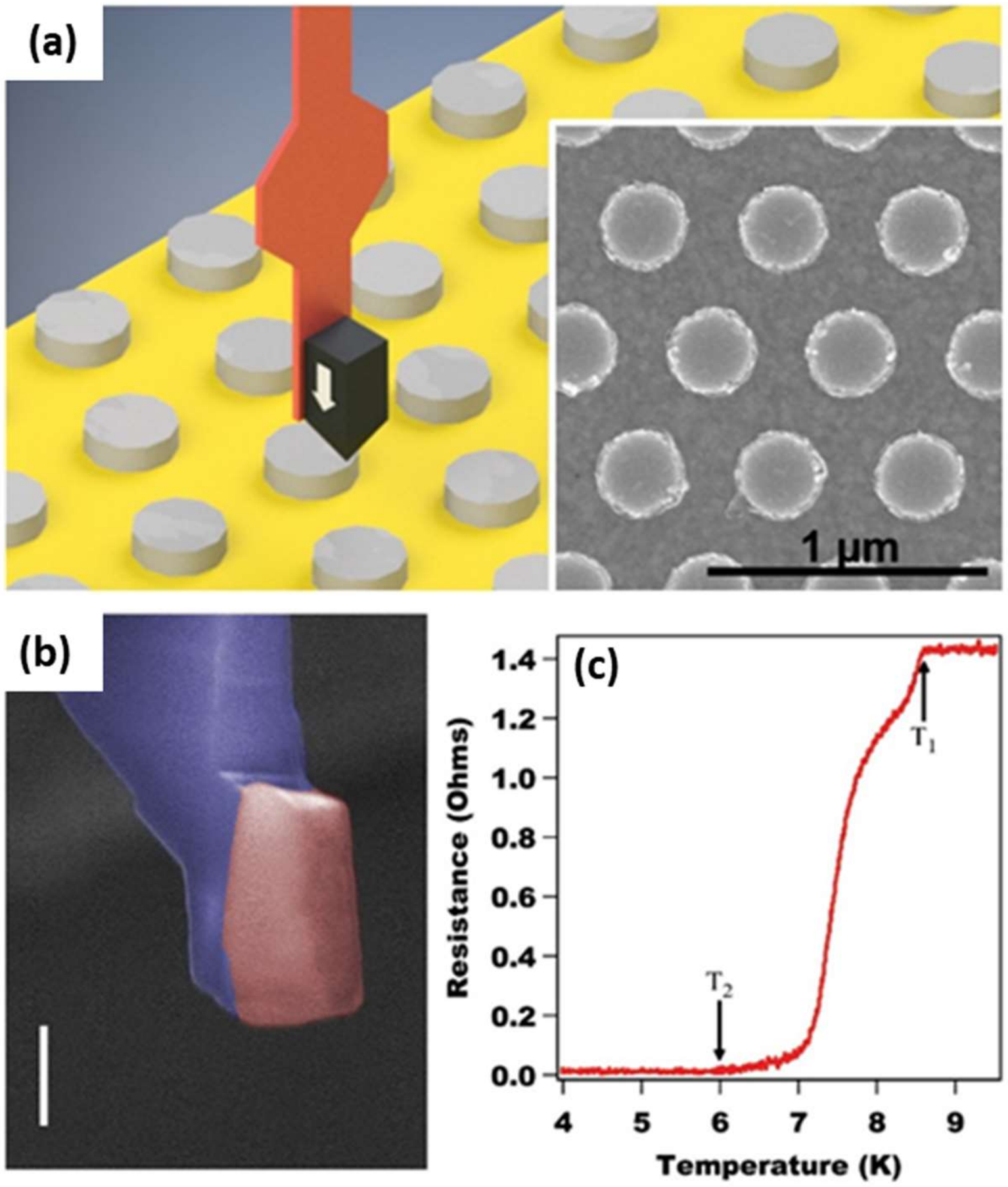

FIG. 1. SNS array and magnetic tip characterization. (a) Schematic of cantilever over a triangular array of Nb islands on top of an Au film. A $SmCo_5$ magnetic tip (black shape) is attached to the end of the cantilever and used to trap vortices. The white arrow on the magnetic tip indicates the direction of the tip's magnetization. Inset: Scanning electron microscope (SEM) image of an array with 500 nm center-to-center spacing. (b) False-color SEM image of one of the $SmCo_5$ magnetic tips used in this work. White scale bar is 500 nm. (c) Temperature dependence of the resistance near the superconducting transition.

for vortices [29–31].

Scanning measurements are performed below $T_2$ using an ultra-soft micromachined Si cantilever, mounted in a pendulum configuration, with an $SmCo_5$ magnetic tip shaped via focused ion beam [Fig. 1(b)]. All MFM scans are taken below $T_2$ to allow us to neglect vortex nucleation due to thermal fluctuations and assume that all phase windings and density suppressions in the superconducting order parameter are due to external or tip magnetic fields. An estimate of the tip field is obtained by imaging flux entry into superconducting Al rings (see Appendix A). A uniform magnetic field applied perpendicular to the SNS array, and anti-parallel to the field of the tip, tunes the number of vortices trapped underneath the cantilever tip. The magnetic tip creates a potential well underneath it for vortices with a particular circulation, and at the same time, it repels oppositely circulating vortices that are generated by the uniform field applied to the SNS array. Hence, underneath the tip several tip-induced vortices will be trapped. The tip field controls both the depth and width of the well while the external applied field affects only the depth of the potential. Overall, the external field serves as an additional tuning parameter which globally defines the vortex array that is then modulated by the tip field. As parameters related to the vortex energies and interactions are changed (e.g., tip height, magnetic field magnitude, array spacing) the configuration of the trapped vortices changes to minimize the local energy. As the vortices move, frequency shifts of the cantilever are generated due to the interaction between the vortex and cantilever fields.

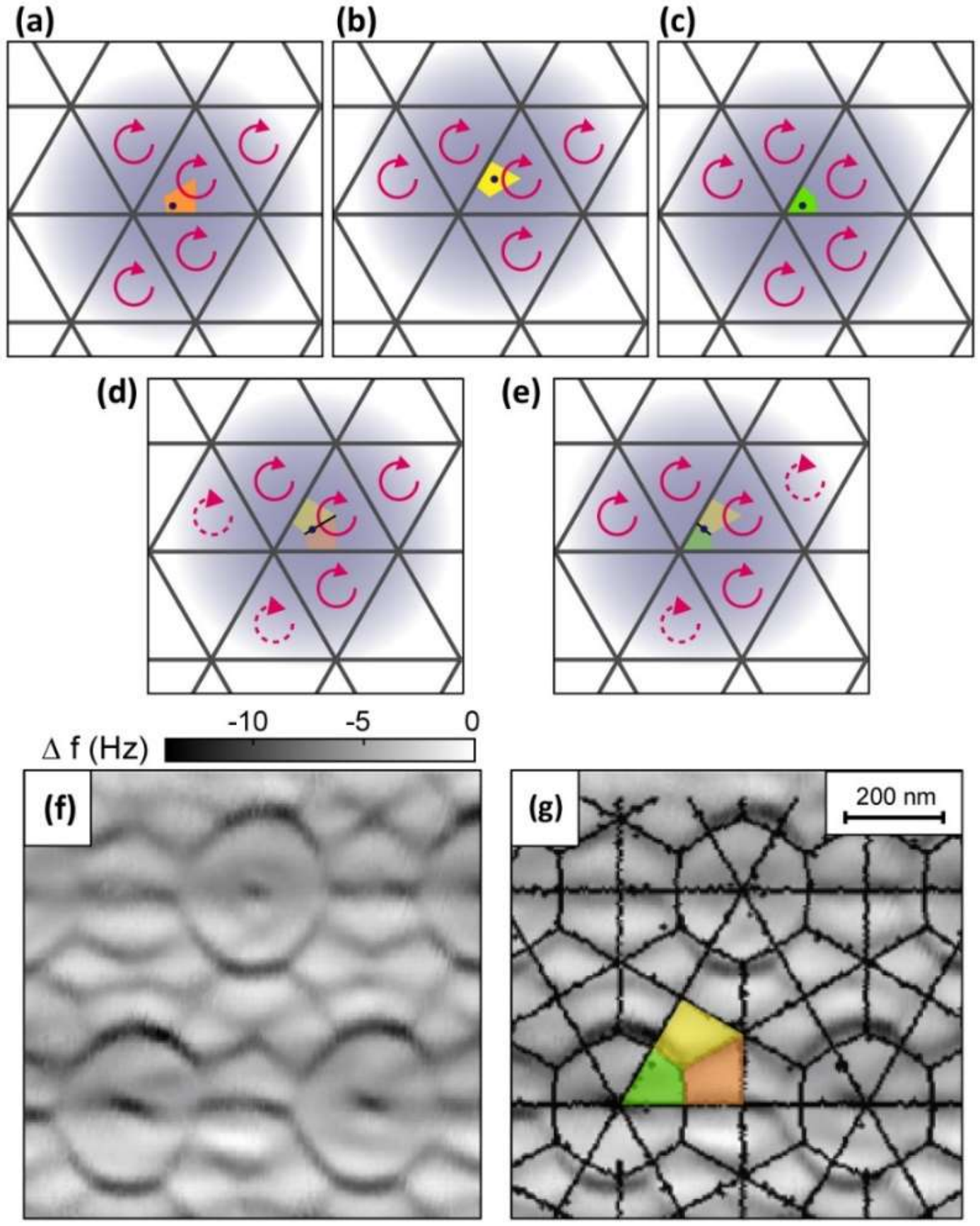

FIG. 2. Schematic showing formation of spatial frequency shift maps. (a)-(c) Several different vortex configurations are stabilized for different positions of the tip with respect to the island array. Here the vertices of triangles correspond to the positions of Nb islands and the shaded blue region marks the field on the surface of the sample with the center indicated by the dot. Vortices in the interstitial regions are shown by red circular arrows. Small colored regions mark the positions of the tip center for which the vortex configurations are stable. (d), (e) When the tip encounters a position where the vortex configuration changes (black line), it will drive the vortices between these configurations as it oscillates, leading to a force on the cantilever and the emergent resonant frequency shift. (f) Representative example of an experimental frequency shift. (g) Simulated locations of the transitions between vortex configurations overlaid with the experimental data. Shaded regions correspond to vortex configurations shown in (a)-(c).

To generate spatial frequency shift maps, the cantilever is raster scanned over the surface of the SNS array at a fixed tip scanning height, with a small fixed oscillation amplitude (typically $\sim$ 15 nm), which perturbs the position of the potential well that traps vortices [Fig. 2(a)-(e)]. The cantilever is kept oscillating at its resonant frequency [32] and is monitored by a phase-locked loop. When the cantilever moves across the array at a fixed applied field the energies of two distinct configurations of vortices can become degenerate at certain tip locations [Fig. 2(d), (e)]. While the cantilever is over these degeneracy locations, the oscillations of the cantilever, along with thermal excitations of the vortices, will drive the vortices between the two configurations in resonance with the cantilever, leading to a force on the cantilever and an associated frequency shift. This frequency shift, which we found to be $\Delta f \approx$ 5-15 Hz for our cantilevers with resonant frequency of $f_0 \sim$ 4 kHz, vanishes quickly as the cantilever moves away, sharply marking the positions where the vortex configurations become degenerate. Raster scanning over the entire imaging window leads to a spatial map of frequency shifts.

In Fig. 2(f) we present an experimental frequency shift map for an island spacing of 500 nm, where dark lines indicate the boundaries between two stable vortex configurations and the lighter areas show tip positions where the vortex configuration is stable and thus there is no associated frequency shift of the tip. The figure shows an intricate pattern of approximately periodic frequency shifts. This pattern consist of a series of nearly circular features, which are centered on the superconducting islands, and a series of lines, some of which are horizontal, and others appear in zig-zag configuration. Once these spatial frequency shift maps are obtained experimentally, we can infer the corresponding vortex configuration from Monte Carlo simulations that minimize the energy of the system (see Section II.A). For example, our modelling indicates that the frequency map shown in Fig. 2(f) corresponds to a 5-vortex configuration. Figure 2(g) shows the comparison of the experimentally observed pattern with the calculated locations of transitions between configurations of 5 vortices.

### A. Numerical simulations of vortex configurations

To understand the underlying mechanism that causes the formation of these patterns, we will now discuss how these images encode information about vortex dynamics in the SNS array. We performed numerical simulations of a simple phenomenological model of vortices. We model the system as an array of Josephson junctions, approximating the Josephson current as $I \approx I_c \gamma_{ij}$, where $\gamma_{ij}$ is the gauge-invariant phase between islands $i$ and $j$. This approximation allows for several convenient simplifications to the effective vortex energy (see Appendix B). We assume that each vortex is point-like and sits in the center of a plaquette, and the subsequent model for the vortex energy is

$$E[n] = p_{\text{int}} \sum_{p,q=1}^{N_{\text{plaq}}} \mathrm{V}_{pq} n_p n_q + \sum_{p=1}^{N_{\text{plaq}}} \left[ \left(U_f\right)_p + \mu_{\text{vort}} \right] n_p \qquad (1)$$

where $N_{plaq}$ is the number of plaquettes, $p_{int}$ is a relative scale factor between the vortex-field ($\left(U_f\right)_p$) energy and vortex-vortex $\left(V_{pq}\right)$interaction term, $\mu_{\text{vort}}$ represents the chemical potential of the vortices, and $n_p$ is the number of vortices in plaquette $p$. We use a classical Metropolis algorithm Monte Carlo simulation to determine the lowest energy vortex configuration for a fixed vortex number (Appendix B). We then compared the lowest energy vortex configurations for differing vortex numbers to determine the configuration with the lowest overall energy, hence identifying the vortex number and its configuration for a given tip location. By tuning the relative strengths of $(U_\mathrm{f})_p$, $V_{pq}$, and $\mu_\mathrm{vort}$, to fit the data at the correct external field and tip heights, we can extract the relative energy scales of the system.

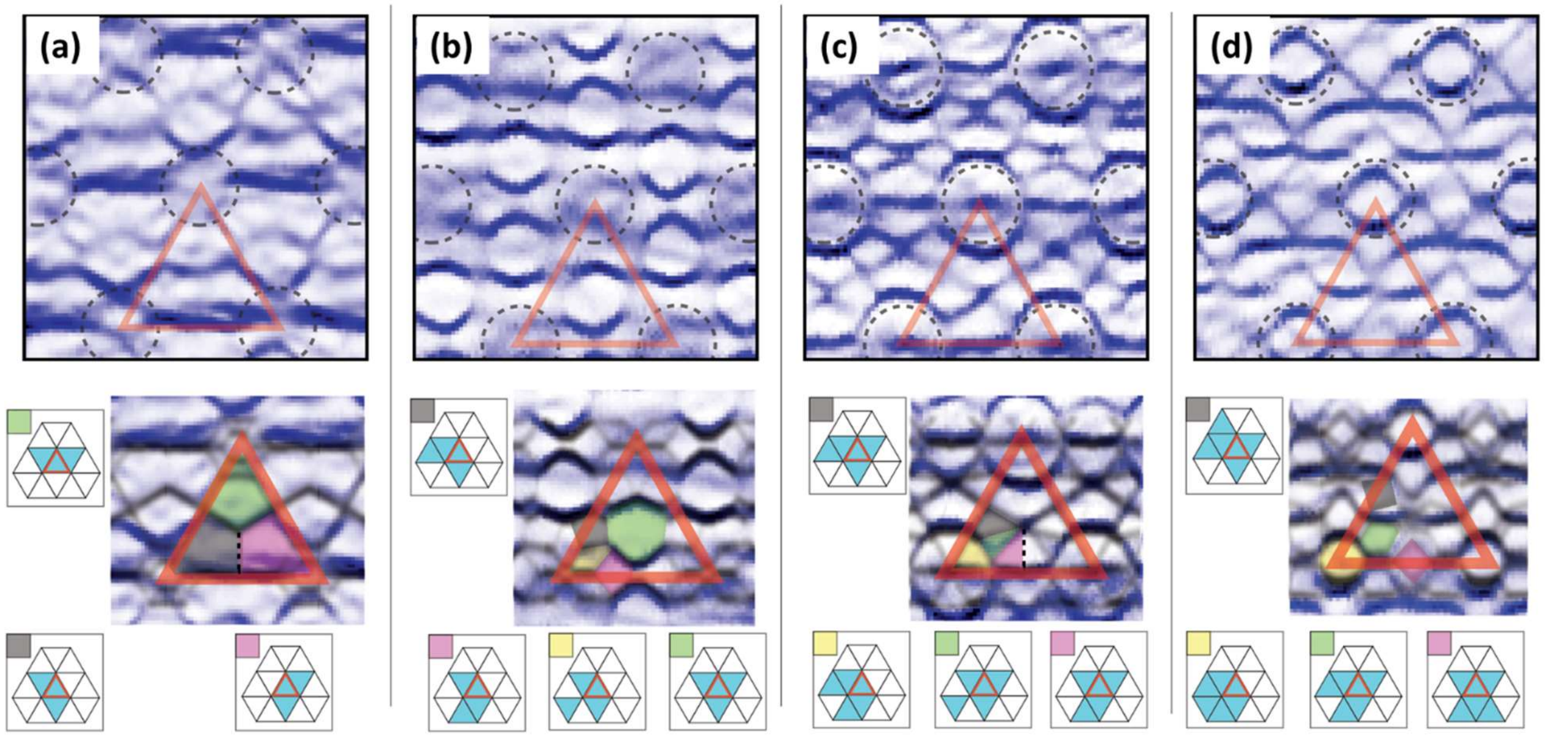

FIG. 3. Images of some patterns seen in this experiment (top) and associated vortex configurations (bottom) as determined by simulated annealing. A slowly varying background was removed from all images to highlight the pertinent features. Configurations are shaded where they are the lowest energy state. The simulation data is darker in areas where the cantilever would experience a larger frequency shift due to the oscillating current. Some dashed vertical lines are added to the simulation to highlight stable regions for a given vortex configuration. One plaquette (red triangle) and associated islands (dashed circles) are drawn for clarity. Experimental and simulations taken for (a) 3 (124 Oe, 350 nm), (b) 4 (85 Oe, 425 nm), (c) 5 (80 Oe, 425 nm), and (d) 6 vortices (68 Oe, 425 nm). All images taken at 3.70 K, except (b), taken at 3.75 K.

As an example of these simulations, in Fig. 3(a) we show the patterns and associated vortex configurations produced by three vortices. As can be seen, there is very good agreement between the simulations obtained from the model we use (bottom) and the experimental measurement (top). By increasing the number of vortices by one and running the simulation again, the resulting pattern obtained changes and reproduces another of the experimental scans, as shown in Fig. 3(b). Using this technique, we can thus show that Figs. 3(a), (b), (c), and (d) demonstrate the energy landscapes and corresponding vortex configurations for 3, 4, 5, and 6 vortices, respectively.

### B. Effects of tip scanning height and applied magnetic field

Figures 4 and 5 show additional real space maps of the frequency shifts associated with changes in vortex configurations, where striking geometric patterns are produced as the applied field, tip height and array spacing are tuned. Changes in the number of vortices and their energy landscapes are manifested as a remarkable evolution between frequency shift patterns. We can controllably alter the number of vortices and thus vortex configurations by tuning three main parameters: the applied magnetic field, tip scanning height, and array spacing.

Figure 4(a)-(e) show frequency shift maps for a 500 nm-spaced island array obtained by fixing the applied magnetic field to 50 Oe and tuning the tip height from 540 nm to 340 nm. Through Monte Carlo simulations, it was determined that Fig. 4(a)-(e) correspond to a 3, 4, 5, 6, 7-vortex configuration, respectively. As the tip scanning height is decreased at a fixed applied field, the depth of the potential well

increases, leading to an increase in the number of vortices trapped underneath the tip. Correspondingly, the energy landscape evolves as the vortex number increases.

The tip heights at which a vortex configuration transition happens (for example a transition from a 3-vortex to a 4-vortex configuration) also depends on the applied magnetic field. Fig. 4(f)-(j) show frequency shift maps obtained by fixing the applied field to a higher value of 80 Oe and tuning the tip height from 380 nm to 240 nm. Similar features are evident in Fig. 4(a)-(e) and Fig. 4(f)-(j); from simulation, it was determined that Fig. 4(f)-(j) also correspond to a 3, 4, 5, 6, 7-vortex configuration. However, for the larger applied field the depth of the potential well trapping the tip-generated vortices is smaller and thus there are fewer trapped vortices. For example, by fixing the tip height to 380 nm and tuning the magnetic field from 50 Oe to 80 Oe, as shown in Fig. 4(d) and Fig. 4(f), the number of trapped vortices decreases from 6 to 3. It is clear that as the applied field increases, lower tip heights are required to achieve the same vortex configuration.

Although the frequency shift maps corresponding to the same vortex configurations are similar, they are not identical. For example, both Fig. 4(d) and Fig. 4(i) describe a 6-vortex configuration, however, they differ in that Fig. 4(i) has more well-defined features and also shows additional features not evident in Fig. 4(d). The differences in the patterns arise from a competition between vortex-tip and vortex-vortex interactions. With the tip closer to the surface of the array, vortex-tip interactions dominate, and the vortices are positioned in neighboring plaquettes. At higher tip heights, vortex-vortex interactions begin to dominate, and the vortices will begin to spread out underneath the tip and thus no longer reside in neighboring plaquettes (see Discussion for further details).

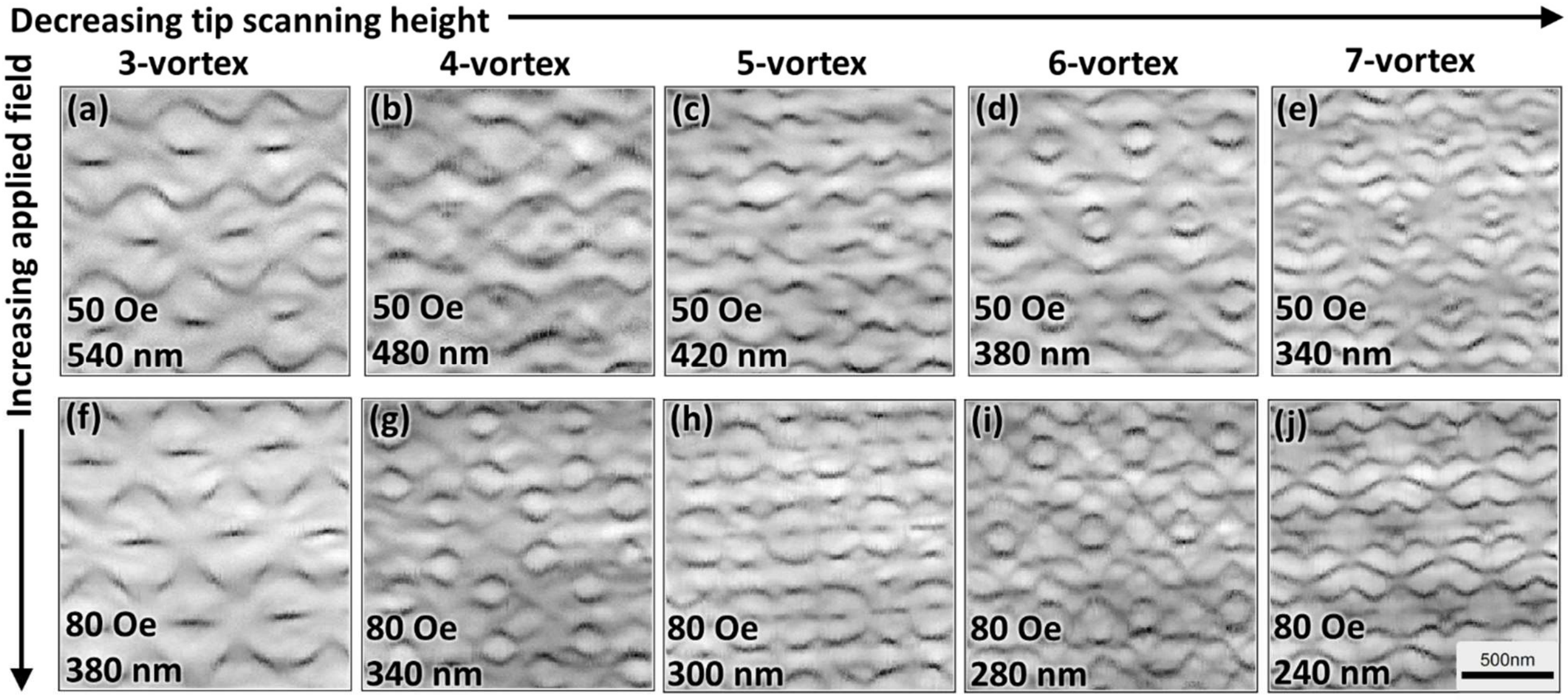

FIG. 4. Vortex configuration changes with tip height and applied magnetic field. Tip height decreases from left to right and applied field increases from top to bottom. The applied field (in Oe) and the tip height (in nm) are indicated on the lower left corner of each panel. (a-e) Applied field is fixed at 50 Oe as the tip height is tuned from 540 to 340 nm. As the tip height decreases, frequency shift maps were taken for (a) 3, (b) 4, (c) 5, (d) 6, and (e) 7 vortices. (f-j) Applied field is increased and fixed at 80 Oe as the tip height is tuned from 380 to 240 nm. As the tip height decreases, frequency shift maps were again taken for (f) 3, (g) 4, (h) 5, (i) 6, and (j) 7 vortices.

Altogether, the tip height and applied field will determine the number and energy landscape of trapped vortices, while the tip position affects the location of the vortices within that configuration. This shows that by constructing an appropriate field profile from the tip, control over the vortex states and extraction of key energy scales can be achieved. Using the model described in Section IIA, we can indeed infer the interaction energy scales by mapping the experimentally obtained images with our simulations.

### C. Effects of array spacing

As the array spacing is tuned, control over the distribution of vortices can also be achieved. In addition to the 500 nm-spaced array discussed in Figs. 2, 3, and 4, we measured arrays with a spacing of 440 nm and 560 nm. Fig. 5(a)-(c) show frequency shift maps of a 6-vortex configuration for each array spacing at approximately the same tip scanning height. The frequency shift map for the closest-packed array with a 440 nm spacing [Fig. 5(a)] was taken at a tip height of 480 nm, an applied field of 35 Oe, and at a temperature of 7.7 K. The frequency shift map for the 500 nm array [Fig. 5(b)] was taken at a tip height of 440 nm, an applied field of 40 Oe, and a temperature of 5.7 K. Lastly, the frequency shift for the 560 nm array [Fig. 5(c)] was taken at a tip height of 440 nm, an applied field of 65 Oe, and at a temperature of 3.85 K. Measurements were made at different temperatures to ensure that all scans were taken at temperatures which are slightly below $T_2$ [Fig. 1(c)]. The $T_2$ value depends on the array spacing [31] where the smaller the array spacing, the larger the value of $T_2$ up to the saturation value of bulk Nb $T_c$; thus the frequency shift maps were taken at the appropriate temperatures right below $T_2$, with the 440 nm array having the largest temperature value and the 560 nm array having the smallest temperature value.

The magnitude of the applied field needed to create a vortex also depends on the array spacing through the magnetic frustration $f = \frac{BA}{\Phi_0}$, where $B$ is the applied field, $A$ is the plaquette area and $\Phi_0$ is the flux quantum. As the plaquette area (proportional to the array spacing) decreases, the field needed to create a vortex in every plaquette ($f$ = 1) increases. Since the applied field is *anti-parallel* to the tip-field, the relation between applied field and number of trapped vortices underneath the tip is reversed – the smallest array spacing requires a smaller applied field to produce the same 6-vortex configuration as the largest array spacing in Fig. 5.

It can be seen in Fig. 5(a)-(c) that there are similarities in the patterns, but they are not identical even though they all represent the energy landscape of 6 vortices. These differences again arise due to vortex-vortex interactions in the system. As the array spacing decreases, the tip field encompasses more plaquettes, enabling the vortex-vortex interaction to spread out the vortices. Overall, the array spacing serves as an additional tuning parameter; as we show in the Discussion, this allows for the extraction of the relative strength of vortex-vortex and vortex-field interactions, which depend on whether the islands are either loosely or closely-packed.

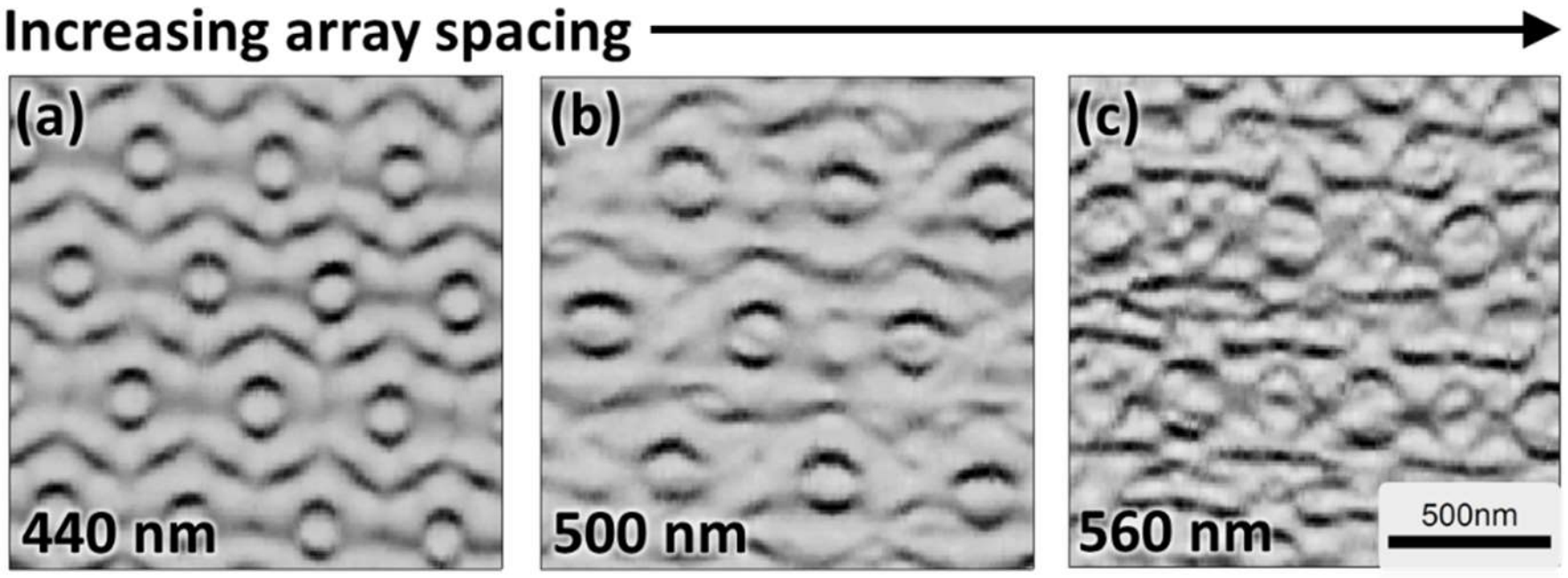

FIG. 5. Vortex pattern changes as a function of array spacing. 6-vortex configuration patterns at a tip height of approximately 480 nm for an array spacing of (a) 440 nm, (b) 500 nm, and (c) 560 nm.

### D. Effects of applied DC driving current

We applied a DC current to the SNS arrays and again performed $\Phi_0$-MFM scans. The application of a DC current results in a spatial shift of the frequency maps perpendicular to the current direction. Fig. 6(a)-(c) show frequency maps for a 6-vortex configuration at a tip height of 270 nm for an applied DC driving current of -35, 5 and 35 μA, respectively. We find that the pattern shown in 6(a)-(c) shifts to the left by about 55 nm as the applied current increases from -35 to +35 μA; the red circles are fixed guides added to emphasize the relative shift of the patterns. We attribute this spatial shift to the Lorentz force acting on the vortices as the current is applied. Fig. 6(d) shows the relation between the applied current and measured spatial shift of the frequency maps for the 500 nm array when the vortex number and tip height is tuned. The shift is determined from the calculated cross-correlation between two images taken with DC currents of opposite directions (see Fig. 6(d) inset).

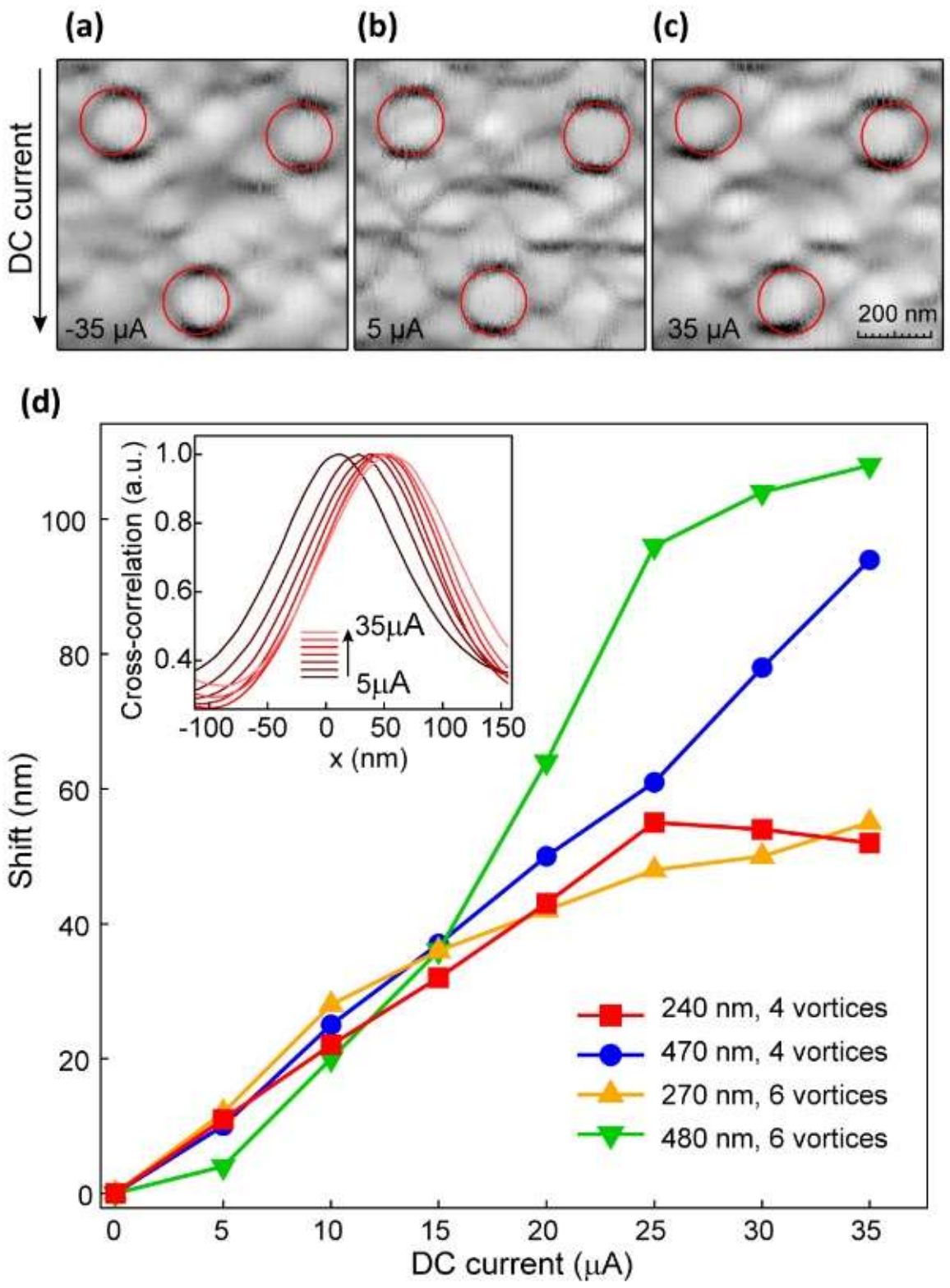

FIG. 6. Frequency maps as a function of an applied DC current. 6-vortex configuration at a tip height of 270 nm with an applied DC current bias of (a) -35 μA, (b) 5 μA, and (c) 35 μA. The red circles are fixed guides added to emphasize the relative shift of the patterns. (d) Spatial shift (in nm) of the frequency maps vs. the applied DC current for a 500 nm array spacing with a 4 and 6-vortex configuration at tip scanning heights of approximately 250 nm and 470 nm. Inset shows the cross-correlation curves used to determine the relative shifts between the images taken for 6-vortex configurations at 270 nm height.

We find that for low applied currents ($I$ < 20 μA) the pattern shift is approximately proportional to the current (Fig. 6(d)). Remarkably, the corresponding coefficient does not show noticeable dependence on the number of vortices or the height of the tip controlling the shape of the confining potential. This behavior suggests that the pinning force, which is a linear restoring force arising when each vortex is shifted from the center of its interstitial region [33], counteracts the Lorentz force due to the transport current. Since the Lorentz force acting on each vortex is $\boldsymbol{F} = \boldsymbol{J} \times \boldsymbol{z} \cdot \Phi_0$ [29], where $\boldsymbol{J}$ is current density, $\boldsymbol{z}$ is a unit vector aligned with the vortex, and $\Phi_0$ is the superconducting flux quantum, we find that the effective spring constant in the linear regime is $k \approx 1.6 \times 10^{-8}$ N/m. It is worth pointing out that the spring constant depends on the relative strength of $J$ and intrinsic supercurrents flowing between the islands which are of the order of $I_c$. Therefore the measurement of $k$ opens a pathway towards extracting the absolute values of vortex energy scales. Such analysis, however, goes beyond the scope of this work and will require calculation of vortex energy with a more realistic model which takes into account the distribution of currents within plaquette and other factors.

In a 4-vortex configuration with a tip height of 470 nm, the shift increases linearly up to the maximum applied current, however at tip height of 240 nm the linear behavior persists only up to about 25 μA [Fig. 6(d)]. At higher currents the shift is independent of the applied current signaling a crossover to a regime in which vortex-vortex and vortex-field interactions likely dominate over the pinning force. The effect of decreasing the tip height while keeping number of vortices constant means that the tip field lines are now confined to fewer plaquettes on the array. This confinement coupled with large driving currents leads to the increased vortex-vortex and vortex-field interactions.

In a 6-vortex configuration at similar scanning heights of 270 nm and 480 nm [yellow and green traces in Fig. 6(d), respectively], the crossover to dominant vortex-vortex and vortex-field interactions happens at lower driving currents when compared to the 4-vortex configuration. More vortices confined over the same number of plaquettes on the array will be more susceptible to crossover to a vortex-vortex and/or vortex-field interaction regime at lower driving currents.

At even higher applied currents, when the Lorentz force dominates over the pinning force, vortex-vortex and vortex-field interactions lead to bulk depinning of the vortices. Once the vortices de-pin from their potential wells, the tip no longer detects frequency shifts and the image is smeared (see Appendix C).

## III. DISCUSSION

We fit patterns at different external fields and tip heights to extract valuable, and previously inaccessible, information about the energy scales that determine vortex dynamics in these systems. Using Eqn. (1), we find that, for a 500 nm center-to-center (inter-island) array, the chemical potential term is approximately $\mu_{\mathrm{vort}} = (1.8 \pm 0.1)\ V_{pp}$, where $V_{pp}$ is the energy of a lone vortex trapped in the array with no fields applied, as presented in the model described in Appendix B, and $p_{\mathrm{int}}$ is approximately $1.0-1.2$. We do not find any dependence of $\mu_{\mathrm{vort}}$ on the number of vortices underneath the tip for the configurations examined. Separate arrays with spacings of 440 and 560 nm were also imaged, and $\mu_{\mathrm{vort}}$ and $p_{\mathrm{int}}$ were extracted. For the 560 nm array, we found $\mu_{\mathrm{vort}} = (0.9 \pm 0.1)\ V_{pp}$, with $p_{\mathrm{int}} \sim 0.7-0.9$, i.e., showing vortex-vortex interactions are weaker relative to vortex-field interactions. We also find that $p_{\mathrm{int}}$ depends on the external field and decreases for higher external magnetic field values. The 440 nm array has $\mu_{\mathrm{vort}} = (2.4 \pm 0.1)\ V_{pp}$, with $p_{\mathrm{int}} \sim\ 1.2-1.4$, indicating stronger vortex-vortex interactions relative to the vortex-field interaction. For this lattice spacing we find that $p_{int}$ increases for higher field values.

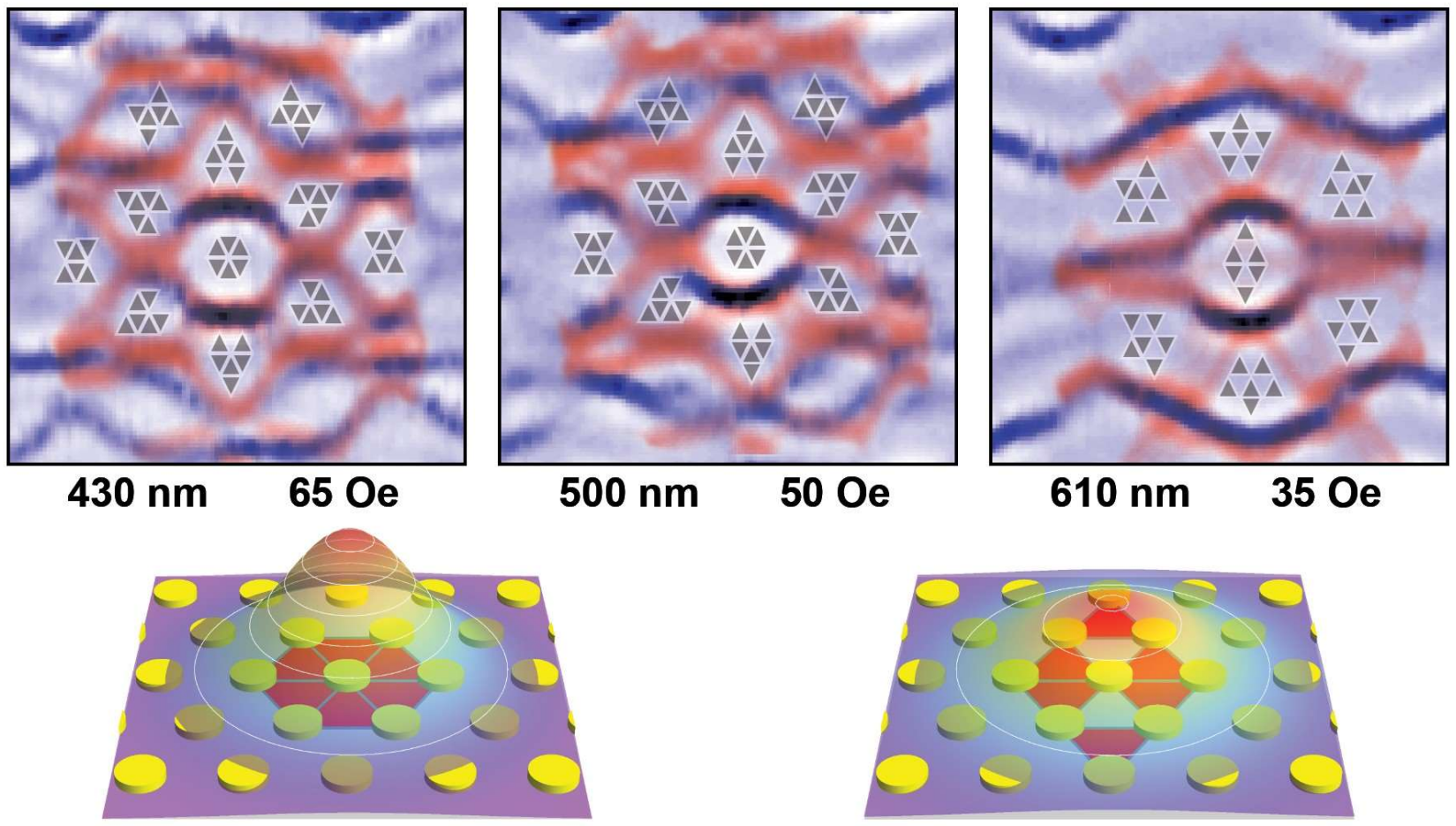

FIG. 7. Simulated vortex pattern changes with height for 6 vortices underneath the tip. Tip height increases from left to right, as shown below each image. Images are overlaid with data from simulations (red) and the stable vortex configuration in each region. As the tip height is increased, the potential well flattens out, while the associated external field is changed to keep the number of vortices constant. Some stable vortex configurations cover a smaller area as the tip height is increased, with those regions disappearing in the furthest image. In the rightmost image, $\mu_{\mathrm{vort}}$ was decreased to $1.4V_{pp}$ to achieve a better fit. The lower images show the field distribution on the surface for 430 nm (left) and 610 nm (right) tip separations with the tip over the central feature of the top images.

In addition to the extraction of these characteristic energy scales, some in-situ control over the vortex configurations is achieved by varying the height of the tip as previously discussed. By simulating the effects of the tip height on a certain vortex configuration, we were able to further extract the location of each vortex within the array. Fig. 7 shows the evolution of configurations of 6 vortices and accompanying simulations as the tip height is increased and the applied field is tuned so to keep the number of vortices constant. At a tip height of 430 nm, it was demonstrated by the simulation that the vortices reside in neighboring plaquettes (Fig. 7). As the tip height is increased, the vortices begin to spread out and begin to reside in next-nearest neighboring plaquettes. Some vortex configurations that are present with a deeper well will cover less area in the image or can even disappear as the well is made shallower. This again is due to the vortex-vortex interactions becoming relatively stronger, and hence more significant in determining the vortex configurations for these conditions.

## IV. CONCLUSIONS

In conclusion, we have demonstrated a robust experimental platform for locally probing and controlling vortex dynamics. By trapping vortices underneath a magnetic tip, we can characterize transitions between stable vortex configurations and are able to extract the relative energy scales of various interactions. We tune the number and distribution of vortices trapped underneath the tip by modifying the scan height, external field, and array spacing. Using simulations of a simple model of vortices, we can reproduce the observed image patterns. The versatility of this experimental platform could prove a powerful tool to obtain a local understanding of, for example, the dominant effects that lead to various forms of vortex matter in superconductors such as vortex glasses and vortex liquids. We note that this technique is useful for mapping and manipulating vortex configurations even in systems without periodic pinning potentials. Furthermore, this technique has the potential of probing non-standard vortex interactions in novel superconducting systems. In particular, these results open interesting opportunities for applications in quantum computing platforms that require the manipulation and braiding of vortices.

## ACKOWLEDGEMENTS

This work was supported by the DOE Basic Energy Sciences under DE-SC0012649 and the National Science Foundation (NSF) under DMR 17-10437. V.C. was supported by the Gordon and Betty Moore Foundation EPiQS Initiative through Grant GBMF4305. NM also acknowledges support from DOE-EFRC under DE-SC0021238 for analysis/manuscript preparation. This research was carried out in part in the Materials Research Laboratory Central Research Facilities, University of Illinois.

## APPENDIX A: MATERIALS AND METHODS

### 1. Device Fabrication

Electron beam lithography and electron beam evaporation were used to define and deposit several layers of material. The first, an 18 nm Au layer, with an underlying 1 nm Ti adhesion layer was placed onto a Si substrate with 300 nm $SiO_2$ as an insulator. The second layer consists of Al registration marks to aid in determining the location of the tip on the surface. A final round of processing was used to define and deposit the Nb islands. Prior to the Nb deposition, the surface of the Au was $Ar^+$ ion milled to establish a clean interface, and the Nb was evaporated at a pressure of $\sim 10^{-9}$ Torr. For one sample, one 500 nm center-to-center spaced array was made on an Au pad of 80 $\mu$m × 80 $\mu$m. For the second sample, two 500 nm spaced arrays were 50 $\mu$m × 50 $\mu$m, with one connected in a four-point configuration. The connected array was used to determine the transition temperatures [Fig. 1(c)] and the magnetoresistance of the 500 nm spaced arrays, while the other was used for imaging experiments. Further arrays on the same sample, with lattice spacings of 440, 500, and 560 nm were also imaged. These arrays had areas of 50 $\mu$m × 15 $\mu$m.

The cantilevers used in this work were custom-fabricated Si cantilevers of length 110 $\mu$m, width 4 $\mu$m, and thickness 100 nm. A $SmCo_5$ magnetic particle was positioned on the end of each cantilever using a micromanipulator, aligned with the cantilever axis using an external magnet, then epoxied into position. The magnet was then shaped using a Ga focused ion beam with low (< 10 pA) current to preserve the magnetization of the $SmCo_5$. Torque cantilever magnetometry was used to measure the magnetic moment of the tip and ensure that it is well aligned with the cantilever axis.

### 2. Measurement

Measurements were taken in a He-3 refrigerator with a base temperature of 300 mK. Cantilever oscillations were measured using a laser interferometer, and the cantilever was self-oscillated using a feedback loop at a small amplitude, typically 15 nm. Frequency was determined using a phase-locked loop running on an FPGA. Images were taken at least 5 $\mu$m from the edge of the arrays to minimize edge effects. Images were raster-scanned using an ANSxyz100 (Attocube) piezoelectric scanner, with the fast axis in the $y$-direction (vertical), at a rate of less than 300 nm/sec.

### 3. Tip Field Estimate

To estimate the tip field, the magnetic tip was scanned over superconducting Al rings deposited via e-beam or thermal evaporation. The rings used had radii of 2-5 $\mu$m, with wall thicknesses of ~ 200 nm. Sufficiently close to the superconducting transition, the fluxoid transitions in the ring become reversible and occur when the tip applies a half-integer number of magnetic flux quanta through the ring. The resulting strong interaction between the magnetic tip and the switching supercurrent shifts the resonant frequency of the cantilever [27, 28]. We mapped the locations of these frequency shifts as positions where the flux through the ring has changed by one flux quantum.

A model of the tip was then created consisting of $50 \times 50 \times 50$ nm$^3$ voxels with a magnetic dipole at the center. The tip magnetization was set to be the measured value, as determined by cantilever magnetometry. A scanning electron micrograph is used to determine where to position the dipoles, and their strength is adjusted to match the observed flux changes as the simulated tip is scanned across a ring. The dipoles are adjusted until the simulated flux changes and observed flux changes line up at multiple scan heights. Estimates of the tip field were then generated from the final dipole configuration. For a detailed description of this procedure for the tip field estimate refer to previous work in Ref. 27.

## APPENDIX B: VORTEX MODEL IN MONTE CARLO SIMULATIONS

The simulated environment consisted of a roughly circular array of plaquettes 50 lattice constants in diameter. Vortices were placed and could move in the central 85 plaquettes of this larger area. Vortices were initialized dependent on the flux in each plaquette from the magnetic tip and external field. Vortex-antivortex pairs were randomly generated in the 85 plaquette area, and the new vortex configuration energy was compared to the previous state. The lowest energy state was selected using a simulated annealing procedure. The minimum energy states generated from this procedure for differing vortex numbers were then compared using the energy formula given in Eqn. (1), and the lowest energy state was selected for each tip position.

### 1. Josephson junction array model

We start with a phenomenological model based on Josephson junction arrays. This amounts to neglecting the SC condensate in the interstitial regions altogether and focuses only on the Nb islands and their inter-island Josephson couplings. The interstitial regions, which host a weaker SC condensate, then act as Josephson weak links. Vortices that occupy the interstitial regions are essentially Josephson vortices in this picture. Thus we consider the following Josephson junction array quantum Hamiltonian [29, 31, 34]

$$\hat{H} = \frac{1}{2}\sum_{i,j} U_{ij}\hat{Q}_i\hat{Q}_j - \sum_{i\neq j} J_{ij}\cos(\hat{\theta}_i - \hat{\theta}_j - \varphi_{ij}[\boldsymbol{A}]) \tag{B1}$$

where the $i, j$ indices label the individual Nb islands. The operators $\hat{Q}_i$ and $\hat{\theta}_i$ refer to the charge 2e Cooper pair number, and SC phase operators respectively. They are mutually conjugate and satisfy the commutation relation

$$[\hat{Q}_i, \hat{\theta}_j] = -i\delta_{ij} \quad \text{(B2)}$$

The first term in $\hat{H}$ is the charging energy with $U_{ij}$ being proportional to the inverse of the capacitance matrix. The second term is the Josephson coupling term with coupling matrix $J_{ij}$ between sites $i, j$. The quantity $\varphi_{ij}[\boldsymbol{A}]$ is an additional phase term that originates from the presence of a magnetic vector potential $\boldsymbol{A}(\mathbf{x})$ associated to a non-zero out of plane magnetic field $B_z$. It ensures that the phase difference

$$\gamma_{ij} = \hat{\theta}_i - \hat{\theta}_j - \varphi_{ij}[\boldsymbol{A}] = -\gamma_{ji} \quad \text{(B3)}$$

on the link between $i$ and $j$ is gauge invariant.

Next, we make three simplifying approximations:

1. The charging term, which is typically small for mesoscopically large SC islands, is discounted. Effectively the Nb islands function as charge reservoirs (Cooper pair boxes) with large capacitances. This turns H into a classical energy functional on the set of island phases $\{\theta_i\}$.
2. The Josephson couplings are limited to only nearest neighbors $(ij)$ of the triangular lattice island array. This is rationalized by the fact that $J_{ij}$ decays with increasing inter-island distance making Cooper pair tunneling between nearest neighbors the dominant interaction. We expect that the reincorporation of the neglected Josephson couplings will not qualitatively change main the results of our analysis.
3. We assume that the value of the phase differences $\gamma_{ij}$ are small, hence legitimizing a Taylor expansion of the cosine. This is equivalent to assuming that the Josephson supercurrents $Iij$ between islands $i$, $j$ are small enough such that $I_{ij} = I_c\sin(\gamma_{ij}) \approx I_c\gamma_{ij}$ , where $I_c$ is the critical supercurrent between nearest neighbors.

With these simplifications, the model is re-interpreted as a *static* Josephson junction array on a triangular lattice with the Josephson supercurrents $I_{ij}$ defined on nearest neighbor links $(ij)$, as the effective degrees of freedom. This has the following effective static energy function

$$E_{\text{eff}}[I] = \frac{E_J}{2I_c^2} \sum_{<i,j>} (I_{ij})^2 \quad \text{(B4)}$$

where $E_J$ is the Josephson energy between nearest neighbor SC islands and we have dropped an irrelevant constant. It is convenient at this point to choose an orientation convention for the links $(ij)$ in organizing the currents $I_{ij}$, and to avoid over-counting. A simple choice is to take a counter-clockwise orientation in the up-pointing triangular plaquettes (Δ) which leads to a clock- wise orientation on the down-pointing triangular plaquettes (∇).

Magnetic flux penetrates the system through the triangular plaquettes of the lattice by an amount $\Phi_{\text{ext}}[p]$ externally applied through plaquette $p$. In the absence of SC vortices, $-\Phi_{\text{ext}}[p]$ is proportional to the supercurrent density circulation $\oint \boldsymbol{j} \cdot \boldsymbol{dl}$ enclosing plaquette $p$ [29] within the Au film. This supercurrent density circulation is proportional to the sum of phase differences which gives

$$-\Phi_{\text{ext}}[p] = \alpha \sum_{i,j\in p}^{\circlearrowleft} I_{ij} \approx \frac{\Phi_0}{2\pi} \sum_{i,j\in p}^{\circlearrowleft} \gamma_{ij}$$

where $p$ is a label of the plaquette, $\Phi_0 = h/2e$ is the flux quanta, and $\alpha > 0$ is a proportionality constant depending on geometry of the system, the magnetic permeability and the condensate density. The sums are taken in the anti-clockwise (↺) sense; for both Δ and ∇ plaquettes. Note that the plaquettes themselves reside in a honeycomb lattice dual to the triangular lattice.

Now, when a SC vortex is present in $p$, the sum of phase differences $\gamma_{ij}$ is of order $2\pi$ and is no longer expected to be small such that the linear approximation $\sin x \approx x$ (assumption (3) above) holds. Nevertheless, we can perform a (large) gauge transformation which changes sum of phase differences by quantized multiples of $2\pi$ or fluxoids

$$\sum_{i,j\in p}^{\circlearrowleft} \gamma_{ij} \to \sum_{i,j\in p}^{\circlearrowleft} \gamma_{ij}\ (\mathrm{mod}\ 2\pi) = \sum_{i,j\in p}^{\circlearrowleft} \gamma_{ij} - 2\pi n_p$$

such that the $\gamma_{ij}$ 's and hence their sum is small once more. Incorporating this into the relation with $\Phi_{\mathrm{ext}}[p]$ yields

$$\sum_{i,j\in p}^{\circlearrowleft} \gamma_{ij} = 2\pi n_p - \frac{2\pi\Phi_{\mathrm{ext}}(p)}{\Phi_0}$$

where $n_p \in \mathrm{Z}$ is an integer that is non-zero whenever a vortex (anti-vortex) is present in $p$. The external flux is more conveniently expressed as

$$\Phi_{\mathrm{ext}}(p) = \Phi_0 f_p \tag{B5}$$

with $f_p$ being the local magnetic flux fraction or frustration at $p$. Thus we have the following constraint equation for each plaquette

$$\frac{1}{I_0}\sum_{i,j\in p}^{\circlearrowleft} I_{ij} = n_p - f_p \tag{B6}$$

where $I_0^{-1} \equiv \alpha/\Phi_0$ is a proportionality constant with dimensions of $[\mathrm{Current}]^{-1}$. A second constraint on $I_{ij}$ is current conservation, or Kirchoff's first law, at each site $i$. We express this as

$$\sum_{j\in\langle ij\rangle}^{\circledast(i)} I_{ij} = 0 \tag{B7}$$

where the symbol $\circledast\ (i)$ denotes the fact that orientation convention of $I_{ij}$ is chosen to be pointing *into* the site $i$. These constraints must hold for all sites $i$. Implicit in these expressions is the neglect of the mutual and self-inductance terms due to the supercurrents themselves which are generally expected to be a small effect [30].

### 2. Counting Independent Currents

Now a unit cell of a triangular lattice has 1 site, 2 plaquettes and 3 links. Hence on average per site, current conservation (Eqn. (B7)) removes 1 independent current/link degree of freedom such that the flux conditions (Eqn. (B6)) relate 2 independent currents $I_{ij}$ to 2 independent vortex numbers $n_p$ given fixed frustrations $f_p$. In the case of a finite lattice with open boundaries, after a proper accounting of the links at the boundary, and noting that there are only ($N_{\mathrm{node}} - 1$) current conservation constraints for $N_{\mathrm{node}}$ sites, we find a 1-1 relation between independent currents and a specified configuration of $n_p$'s on each plaquette. This reduction of the current conservation constraints by one comes from the fact that the entire system must have a net zero current.

This can also been seen by noting the Euler characteristic $\chi = 1$ for a finite planar graph relates $N_{\mathrm{node}} - N_{\mathrm{link}} + N_{\mathrm{plaq}} = 1$ where $N_{\mathrm{node}}$ is the number of island sites, $N_{\mathrm{link}}$ is the number of nearest neighbor links, and $N_{\mathrm{plaq}}$ the number of triangular plaquettes. By rearranging we have $N_{\mathrm{plaq}} = N_{\mathrm{link}} - (N_{\mathrm{node}} - 1)$ which says that $N_{\mathrm{plaq}}$ is the same as the number of independent current links. Hence, for fixed frustrations $\{f_p\}$, specifying a configuration of vortex numbers $n_p$ for all plaquettes is equivalent to specifying a current configuration $I_{ij}$ on all links that obey the required constraints.

### 3. Transforming Currents to Vortex Occupations

By combining the constraints in equations (A6) and (A7), we can relate a configuration of vortex numbers $\mathbf{n} = \{n_p\} \in Z^{N_{\text{plaq}}}$ to a configuration of currents $\boldsymbol{I} = \{I_{ij}\} \in R^{N_{\text{link}}}$ . This relation is linear and can be succinctly expressed as

$$\frac{1}{I_0}\mathrm{M}\mathbf{I} = \begin{pmatrix} \mathbf{n}-\mathbf{f} \\ \mathbf{0}_{N_{\text{node}}-1} \end{pmatrix} \tag{B8}$$

where $\mathbf{f} = \{f_p\} \in R^{N_{\text{plaq}}}$ are the externally applied flux fractions/frustrations and $\mathbf{0}_{N_{\text{node}}-1}$ is a ($N_{\text{node}} - 1$) dimensional zero vector. The matrix M is $N_{\text{link}} \times N_{\text{link}}$, dimensionless and invertible due to the counting arguments just mentioned. Taking the inverse yields

$$\mathbf{I} = I_o\mathrm{M}^{-1}\begin{pmatrix} \mathbf{n}-\mathbf{f} \\ \mathbf{0}_{N_{\text{node}}-1} \end{pmatrix} \tag{A9}$$

### 4. Effective Energy Function

Next, inserting the expression (9) into the effective energy function (4) gives

$$E_{\text{eff}}[\mathbf{n},\mathbf{f}] = \frac{E_j}{2(I_c/I_0)^2}(\mathbf{n}-\mathbf{f})^T B^T B(\mathbf{n}-\mathbf{f}) \tag{B10}$$

where $B$ is a submatrix of $\mathrm{M}^{-1}$ in its first $N_{\text{plaq}}$ columns. The prefactor on the RHS sets the overall energy scale, and the dimensionless constant $I_c/I_0$ encodes geometric information about the lattice. We consider the configuration of local frustrations **f** to be a fixed external knob, and the vortex numbers **n** as variational parameters that are required to minimize $E_{\text{eff}}$. The frustrations **f** are determined by the total amount of flux through each plaquette, and are set by the $B_z$ profile induced by the magnetic tip, and the additional uniform field that moderates the tip field. While calculating **f** we ignored the redistribution of the magnetic flux due to field screening by Nb islands and inter-island supercurrents.

Then by scaling away the overall energy scale, expanding the brackets, and dropping an irrelevant constant we find the following model energy function dependent on **n** and **f**

$$E[\mathbf{n},\mathbf{f}] = \mathbf{n}^T\mathrm{V}\mathbf{n} + \mathrm{U}_{\mathrm{f}}^T\mathbf{n} \tag{B11}$$

where

$$\mathrm{V} = B^T B \tag{B12}$$

$$\mathrm{U}_{\mathrm{f}} = -2B^T B\mathbf{f} = -2\mathrm{V}\mathbf{f} \tag{B13}$$

Note that $E$[**n**, **f**] is determined entirely by the geometry of the lattice and the local flux and current constraints. The first term in $E$[**n**, **f**] represents vortex-vortex interactions while the second is the vortex-field interaction. We note that the matrix V is symmetric and is dense in its off-diagonals. This results in long-range, pair-wise interactions between vortices and externally applied fields. For our simulations we calculate the entries of V for a roughly circular array of triangular plaquettes with a diameter of 50 lattice constants.

The entries in **n** are integers, and due to the large number of possible combinations of vortex positions in **n**, the minimization of $E$[**n**, **f**] is done variationally using Metropolis Monte-Carlo. In practice $n_p$ takes 0, 1 values indicating the absence or presence of a single vortex. We simulate an area consisting of 85 plaquettes in a roughly circular region at the center of the larger, 50 lattice constant array to determine the vortex configurations. Also the total vortex number $n_p$ is varied during the search for the energy minimum, but remains fixed during a single Monte-Carlo run. Since the number of vortices is small ($n_p < 10$) we found that Metropolis Monte-Carlo is converging well to the ground state configuration of vortices.

## 5. Further Phenomenological Fitting

As it stands, there are no fitting parameters in the model, which itself depends heavily on all the assumptions previously discussed. However, to fit to the observed data, we have found it necessary to modify the above form of $E[\mathbf{n}, \mathbf{f}]$. The modified model energy function that we optimize is

$$E[\mathbf{n},\mathbf{f}] = p_{\text{int}}\mathbf{n}^T \mathrm{V}\mathbf{n} + \mathrm{U}_{\mathbf{f}}^T\mathbf{n} + \mu_{\text{vort}}\sum_{p} n_p = p_{\text{int}}\sum_{p,q=1}^{N_{\text{plaq}}} \mathrm{V}_{pq}n_p n_q + \sum_{p=1}^{N_{\text{plaq}}}\left[(U_{\text{f}})_p + \mu_{\text{vort}}\right]n_p \tag{B14}$$

where $p_{\text{int}}$ and $\mu_{\text{vort}}$ are the two phenomenologically introduced parameters. The quantity $p_{\text{int}}$ modulates the relative strength between vortex-vortex to vortex-field interactions. While $\mu_{\text{vort}}$ is a chemical potential for the vortices that is added to fine-tune the favored number of vortices and adjusted so that vortex transitions occur at the observed field/heights in the experiment. These two fitting parameters can be thought of as modifications needed to compensate for the limitations of the assumptions and approximations made. For example, the fixed vortex number during a Monte-Carlo run excludes the possibility of fluctuating vortex numbers during a raster scan of the magnetic tip positions. In general, $\mu$ and $p_{int}$ are not universal and depend on the parameters of the array that control the strength and geometry of supercurrents in system.

From fitting the data, we find the chemical potential $\mu_{\text{vort}}$ of the 500 nm array, to be approximately $(1.8 \pm 0.1) \times \mathrm{V}_{pp}$, where $\mathrm{V}_{pp}$ is the on-site vortex energy. For the 440 nm and 560 nm arrays, $\mu_{\text{vort}}$ is approximately $(2.4 \pm 0.1) \times \mathrm{V}_{pp}$ and $(0.9 \pm 0.1) \times \mathrm{V}_{pp}$, respectively. In this case, $p_{\text{int}}$ is approximately (1.0-1.2) for the 500 nm array, with higher values (1.2-1.4) for the 440 nm array, and lower values (0.7-0.9) for the 560 nm array. These values are dependent on field for the 440 nm and 560 nm arrays. Some uncertainty exists in these values, due to small changes not affecting the patterns generated significantly, as well as possible errors in tip field estimates.

## Appendix C: HIGH APPLIED DC CURRENTS

As described on the main text, the application of a DC current leads to a shift (in nm) of the frequency maps. The shift is attributed to the Lorentz force on the vortices as the current is applied. The forces at play while the current is applied include the Lorentz force, the pinning force, vortex-vortex interaction and vortex-field interactions. At low currents, the pinning force counteracts the Lorentz force and the shift is linear. As the current increases, vortex-vortex and vortex-field interactions likely begin to dominate and the shift becomes independent of the applied current. At high applied currents, the Lorentz force is enough to start bulk de-pinning vortices and the frequency map becomes smeared. Fig. 8 shows a set of frequency maps for an array with 500 nm spacing in a 4-vortex configuration at a tip scanning height of 560 nm. As can be seen from the figure, at a current of 120 μA the frequency map is completely smeared signaling that the vortices are now fully de-pinned from the array.

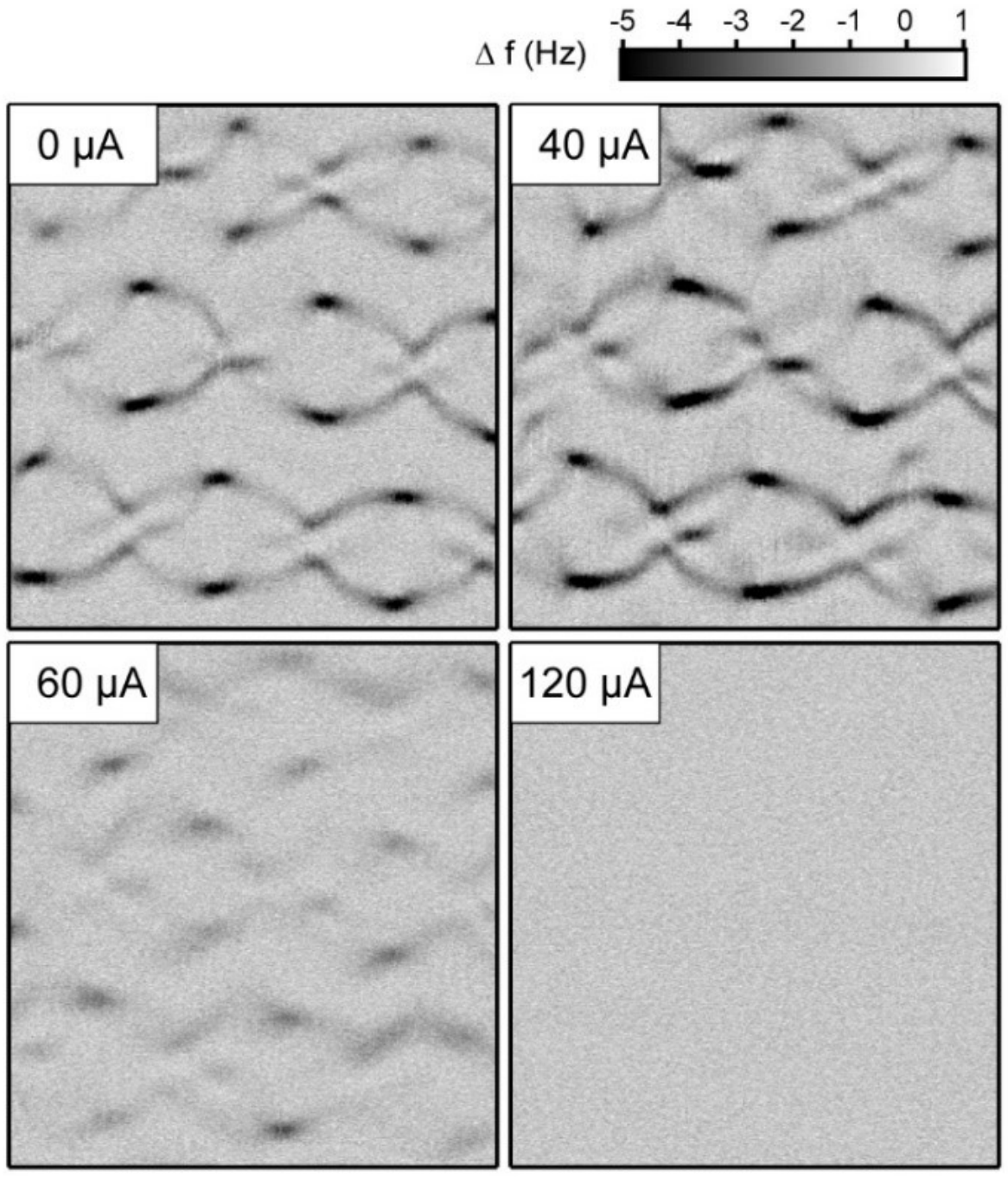

FIG 8: Frequency maps for a 4-vortex configuration in a 500 nm spaced array at a scanning height of 560 nm with an applied DC current of 0, 40, 60 , and 120 μA.